\documentstyle[12pt]{JHEP}

\def\appendix#1{
  \addtocounter{section}{1}
  \setcounter{equation}{0}
  \renewcommand{\thesection}{\Alph{section}}
  \section*{Appendix}
  \addcontentsline{toc}{section}{Appendix \thesection\ \ \ #1}
  }

\newcommand{\newsection}{    
\setcounter{equation}{0}
\section}

\def\slap#1#2{\setbox0=\hbox{$#1{#2}$}
        #2\kern-\wd0{\hbox to\wd0{\hfil$#1{/}$\hfil}}}
\def\sla#1{\mathpalette\slap{#1}}                

\def\on#1#2{{\buildrel{\mkern2.5mu#1\mkern-2.5mu}\over{#2}}}

\def \ov {\over }
\def\bea{\begin{eqnarray}}
\def\eea{\end{eqnarray}}

\def\be{\begin{equation}}
\def\ee{\end{equation}}
\def\ba{\begin{eqnarray}}
\def\ea{\end{eqnarray}}

\def \bi{\bibitem}

\def\id{\protect{{1 \kern-.28em {\rm l}}}}

\title{\LARGE  Superstrings on $AdS_5\times S^5$ supertwistor space }

\author{R.~Roiban and  W.~Siegel\\
C.N.~Yang Institute for Theoretical Physics\\
SUNY at Stony Brook, NY11794-3840, USA\\
Email:roiban@insti.physics.sunysb.edu and 
siegel@insti.physics.sunysb.edu}

\abstract{We derive the Green-Schwarz action on  $AdS_5\times S^5$ 
using an alternate version of 
the coset superspace construction. By Wick rotations and Lie algebra 
identifications
we bring the coset to $GL(4|4)/(Sp(4)\otimes GL(1))^2$, which allows 
us to represent the 
conformal transformations on unconstrained matrices. The derivation is 
more streamlined 
even for the bosonic sector, and conformal symmetry is manifest 
at every step. $\kappa$-symmetry 
gauge fixing is more transparent.}


\preprint{YITP-SB-00-64}

\begin{document}

\newsection{Introduction}

The conjectured duality between $D=4$, ${\cal N}=4$ super Yang-Mills 
theory and Type IIB string theory compactified on five-dimensional 
anti-de-Sitter space
$\times$ the five-sphere ($AdS_5\times S^5$)  \cite{malda}
has created a considerable amount of interest in studying the 
corresponding 
string theory using world-sheet methods. Since the $AdS_5\times S^5$ 
geometry is 
supported by the self-dual Ramond-Ramond (R-R) 5-form background, 
the standard NSR formalism does not apply in a straightforward 
way, while the manifestly 
space-time supersymmetric Green-Schwarz (GS) formalism seems 
more adequate.

In \cite{exactness}, based on maximal supersymmetry arguments, it has 
been shown
that $AdS_5\times S^5$,  $AdS_7\times S^4$ and $AdS_4\times S^7$  are 
exact string 
and M-theory backgrounds.
This raises the hope that in the case of $AdS_5\times S^5$ one might 
find an exact solution of the 
corresponding conformal field theory. The first steps are, of course, writing 
down the world-sheet 
action and fixing its local symmetries.

The GS action in general supergravity background was written 
down in \cite{IIA} for
the type IIA theory and in \cite{IIB} for the type IIB theory. These actions 
are written in terms of 
the space-time vielbein and a super two-form which have as lowest 
components the zehnbein 
and the Neveu-Schwarz-Neveu-Schwarz (NS-NS) two-form. The 
R-R field-strengths and 
the space-time spinors appear as 
higher components. For this reason these actions are not very practical, 
because the full 
solution (to all orders in superspace coordinates) of the supergravity 
constraints is 
not known. 

An approach to constructing the GS action that circumvents this 
problem is the coset (super)space
approach. This requires first that the bosonic background be a 
coset manifold, $G/H$, and furthermore 
that $G$ be the even part of a supergroup. For  $AdS_5\times S^5$ this 
approach was considered
in \cite{coset} and extended to other $AdS_n\times S^n$ 
in \cite{extensions}. In reference
\cite{coset} the authors used the exponential parametrization 
for the coset elements and 
solved the Maurer-Cartan equations. The resulting action can be 
shown to possess $\kappa$-invariance. 
A slight disadvantage of this construction is that the auxiliary 
integral in the Wess-Zumino term 
cannot be performed explicitly without $\kappa$-gauge fixing.  

Yet a third approach to constructing the GS action on $AdS_5\times S^5$ 
was taken in \cite{pes}. This approach circumvents the 3-dimensional 
integral of the Wess-Zumino term
by finding the $AdS_5\times S^5$ potentials already in $\kappa$ 
gauge-fixed form. The resulting 
potentials are then used in an action of the type \cite{IIB}. The resulting 
action was shown to be equivalent to the one obtained in the coset 
superspace approach \cite{agree}.

The action resulting from the first two approaches has to be 
further $\kappa$-gauge fixed. This 
can be done in various ways [9-12], the result being a 
dramatic simplification of the action. The quantization
is nevertheless still problematic since the action also contains 
quartic terms. 

Here we use a version of the coset superspace approach. The main 
difference in our construction is that the space-time super-coordinates 
themselves are a representation of the
superconformal group, not just a nonlinear realization. This implies that 
in the resulting 
action the superconformal symmetry
is manifest. By Wick rotations and Lie-algebra identifications we 
reexpress the superconformal group as $GL(4|4)$. We will therefore 
deal with unconstrained matrices. This 
eliminates the need  of exponential parametrization of the coset 
representatives and dramatically 
simplifies the evaluation of the action. The proof of $\kappa$-invariance
is immediate and requires the use of only a subset of the Maurer-Cartan 
equations. The flat space limit
reveals that the coordinates are in the chiral representation of the 
supersymmetry algebra. Fixing 
$\kappa$-symmetry is more transparent. 

In the following section we describe the manipulations needed to 
represent the superconformal 
transformations as unconstrained matrices and construct a new 
supercoset with $AdS_5\times S_5$ as
bosonic part. In the next section we describe two possible gauge 
fixings of the local symmetries of 
the coset construction
which lead to bosonic sigma models with target spaces Wick-rotated  
forms of particular metrics 
on $AdS_5$ and $S^5$. In section 4 we complete the supercoset 
construction and prove that the 
action is $\kappa$-symmetric. Section 5 is devoted to constructing 
the flat space limit of our construction
and showing by explicit computation that it is the standard GS action 
written in $SO(5)\otimes SO(5)$
spinor notation. Sections 6 and 7 are devoted to $\kappa$ gauge fixing. In 
section 6 we recover 
the Kallosh-Rahmfeld-Pesando gauge by requiring that the action be real 
upon undoing the Wick-rotations.
In section 7 we relax the reality condition and construct a simpler 
action, which has only quadratic terms in fermions.

\newsection{Coset construction}

The coset superspace construction of \cite{coset} considers strings 
propagating on the superspace
${PSU(2,2|4)/ SO(4,1)\otimes SO(5)}$, which has as even part 
the $AdS_5\times S^5$ geometry.
The exponential parametrization of the coset representatives is 
quite complicated and the explicit solution for the coset vielbein does 
not allow a transparent gauge fixing.

Starting from these observations we  simplify the starting point of 
the construction.
We perform  Wick rotations and using the Lie-algebra 
identifications $SO(3,3)\simeq
SL(4)$ and $SO(3,2)\simeq Sp(4)$ we obtain
\be
{PSU(2,2|4)\ov SO(4,1)\otimes SO(5)}\rightarrow {PSL(4|4)\ov 
Sp(4)\otimes Sp(4)}~.
\ee
$PSL(4|4)$, just as $PSU(2,2|4)$, is in fact already a coset; it is 
the coset of $SL(4|4)$
by the $GL(1)$ group of elements with superdeterminant trivially 
equal to unity (matrices 
proportional to the identity). Since $PSL(4|4)$ does not have a 
representation in $Mat(4|4)$
but we would like to use a matrix representation, we further 
relax both the $P$ and 
the $S$ constraints by introducing additional scaling factors 
\be
{PSL(4|4)\ov (Sp(4))^2 }\rightarrow 
{GL(4|4)\ov (Sp(4)\otimes GL(1))^2}~.
\ee
where the two $GL(1)$'s can be chosen to act separately on the 
upper-left and lower-right blocks.
One may notice that only positive determinants are generated 
in this way. This slight shortcoming
is fixed at the end, along with Wick rotating back.
In this last formulation we have the advantage of using the 
unconstrained matrices 
of the general linear group and thus the exponential parametrization 
for the coset elements is
no longer necessary. The coset elements ${Z_M}^A$ transform in the 
defining representation 
of the superconformal group, being therefore supertwistor-like objects.
The index $M$ is acted upon by the superconformal group while 
the $(Sp(4)\otimes GL(1))^2$ acts on 
the $A$ indices.

These objects are not really supertwistors because the grading properties 
of $GL(4|4)$ are different from those of the usual 
supergroups. This prevents us from replacing $GL(4|4)$ spinors 
with ``vectors'' as
\be
w^{MN}=Z^{AM}{Z_A}^N
\label{eq:wnogood}
\ee
where ${Z_A}^N$ is the inverse of ${Z_N}^A$.
However, we can still interpret the bosonic coordinates as bilinears 
in the inverse of the coset elements, in the usual way:
\be
w^{mn}=Z^{am}{Z_a}^n~~(\,AdS_5\,)~~~~~~~~~~~~~
w^{{\bar m}{\bar n}}=Z^{{\bar a}{\bar m}}{Z_{\bar a}}^{\bar n}~~
(\,S^5\,)~~.
\label{eq:w}
\ee
The $GL(1)\otimes GL(1)$ factor in the coset can further be used 
to fix one degree of freedom
for both $w^{mn}$ and $w^{{\bar m}{\bar n}}$ separately. This 
interpretation of  $w^{mn}$ and $w^{{\bar m}{\bar n}}$ is indeed 
correct since any 5-vector can be written in 
this form as the following counting of components 
shows: $4\times 4 - (4\times 5)/2 - 1 = 5$ where the second number is 
the dimension of $Sp(4)$. In this language the superconformal symmetry 
of the background 
is not manifest any more; only the conformal and R symmetry are linearly 
realized.  As will be shown 
later, this is nothing but a coordinate transformation 
from conformally flat metrics (in terms of $Z$) to other type of 
metrics. Another advantage of 
this construction is that it eliminates the use of gamma matrices and 
related identities, since the 
coordinates naturally appear as carrying spinor indices.

Similar ``Wick-rotated'' cosets exist also for $AdS_3\times S^3$ and 
$AdS_2\times S^2$. They are
$GL(2|2)\otimes GL(2|2)/(Sp(2)\otimes GL(1)\otimes GL(1))^2$ and 
$GL(2|2)/(GL(1))^4$, respectively. 
In the following we will concentrate on the $AdS_5 \times S^5$ 
construction.

\newsection{Bosonic \label{bosonic}}

The even part of the supercoset we constructed,  which 
is  $[GL(4)/(Sp(4)\otimes GL(1))]^2$,
should produce a sigma model with target space $AdS_5\times S^5$. Since 
it is ``a perfect square'', 
we will discuss one of the two $GL(4)/(Sp(4)\otimes GL(1))$ factors. The 
gauge fixing can 
be done in a variety of ways. There are, however, 
two extreme possibilities and these will lead to the standard conformally flat
$AdS_5$ and $S^5$ metrics, respectively.
For notational convenience we will break the $GL(4)$ matrices 
into  $2\times 2$ blocks.  The global, $GL(4)$ indices will be split 
as $m\rightarrow (\mu,\,\mu')$ while the
local $Sp(4)$ indices will be split as $a\rightarrow (\alpha,\,\alpha')$.
In general, the local $Sp(4)$ transformations
can be used to make diagonal blocks proportional to the identity matrix 
and to relate the off-diagonal blocks. The $GL(1)$ transformations can 
further be used to 
pull out scales. The action for each of the two factors in the bosonic part 
of the coset is just
\be
S=\int\, j^{\langle ab\rangle }\wedge*j_{\langle ab\rangle }
\label{eq:bosaction}
\ee
where $j^{\langle ab\rangle }$ is the coset part 
of $j^{ab}=z^{am}\,{dz_m}^b,~{\bf z}\in GL(4)$, i.e.~the 
antisymmetric $\Omega$-traceless part of $j^{ab}$. As a matter 
of notation, we will use bold-faced 
letters to denote a matrix as a whole as well as its associated vector. The 
antisymmetrization, tracelessness, 
index contraction and inverse of $\Omega$ are defined as
\bea
A_{[a}B_{b]}={1\ov 2}(A_aB_b-A_bB_a)~~\!&{}&\!~~
A_{\langle a}B_{b\rangle }=A_{[a}B_{b]} +
{1\ov 4}\Omega_{ab}A^cB_c\nonumber\\
A^cB_c=\Omega^{ab}A_b B_a~~\! &{}&   \!~~
\Omega^{ab}\Omega_{ac}=\delta^b_c.
\eea

The $AdS_5$ sigma model in the upper half-space form emerges by 
picking a ``triangular gauge'', i.e.~we 
use the $Sp(4)$ invariance
to set the coset representative in triangular form with the upper-left and 
lower-right $2\times 2$ blocks 
proportional to the identity matrix. We further use the $GL(1)$ invariance 
to scale to unity  the upper-left block. Thus,  the coset representative 
in $GL(4)/Sp(4)\otimes GL(1)$
has the form:
\be
{z_m}^a = \pmatrix{{\bf I}& 0\cr {\bf x}& x^0\,{\bf I}}
\label{eq:adsgauge}
\ee 

The current, and its antisymmetric $\Omega$-traceless part, are 
given by:
\be
{j_a}^b=\pmatrix{0 & 0\cr {d{\bf x}\ov x^0}& {dx^0\ov x^0}\,{\bf I}}~~
~~~~~
j^{\langle ab\rangle }={1\ov 2}\pmatrix{-{dx^0\ov x^0}\,\omega & 
{1\ov x^0}\,d{\bf x}^T\,\omega\cr 
{1\ov x^0} \,\omega\,d{\bf x}& {dx^0\ov x^0}\,\omega}
\label{eq:jads}
\ee
where we chose the $Sp(4)$ metric to 
be $\Omega=\left({\omega\atop 0} {0\atop\omega}\right)$.
Using these expressions  we immediately find that the 
action (\ref{eq:bosaction}) 
gives the $AdS_5$ sigma model in the Wick-rotated Lobachevski 
upper half-space:
\be
{\cal L}_{AdS_5} = \left({dx^0\ov x^0}\right)^2-
{dx^{\alpha'\beta}dx_{\alpha'\beta}\ov 2(x^0)^2}
={(dx^0)^2-(dx^1)^2-(dx^3)^2+(dx^2)^2+(dx^4)^2\ov (x^0)^2}~.
\ee
Undoing the Wick rotation that took us from 
$SO(4,2)$ to $SL(4)$ amounts to changing the sign 
of $(dx^3)^2$. If we further Wick-rotate to Euclidean signature we obtain 
the standard Lobachevski 
upper half-space form of the metric.

The other extreme possibility is to first use the $Sp(4)$ symmetry and set 
the coset representative in the antisymmetric form:
\be
{{{z}_{m}}}^{a} = 
\pmatrix{x_+ {\bf I}& -{\bf x}^T \cr {\bf x}& x_- {\bf I}}~~
{\rm with }~~
x_\pm={1\ov 2}(x^6\pm x^0)
\label{eq:s5gauge}
\ee 
This gauge produces, as we will see, the standard conformally flat 
metric on $S^5$, but works equally
well for $AdS_5$. The antisymmetric traceless part of the current has 
the  expression
\be
j^{\langle {a}{b}\rangle }={1\ov 2\|{\bf z}\|^2}
\pmatrix{ (x_+\,dx_--x_-\,dx_+)\omega^{{\alpha}{\beta}}  & 
x^{{\beta}' \alpha}\,d x^6 -x^6\,dx^{{\beta}'\alpha }\cr
          x^6\,dx^{{\alpha}'{\beta}}- x^{{\alpha}'{\beta}}\,
d x^6   
&  (x_-\,dx_+ - x_+\,dx_-)
\omega^{{\alpha}'{\beta}'} }
\ee
with $\|{\bf z}\|^2 = x_+\,x_--{1\ov 2} \,
x^{\alpha'\beta}x_{\alpha'\beta}$ and $x^{\alpha\beta'}=-
x^{\beta'\alpha}$. This expression suggests that it is convenient 
to fix the $GL(1)$ gauge
by requiring $x^6=R$. Defining the object $z=(x^0,\,x^{\alpha'\beta})$
and its square as
$z^2=-(x^0)^2 -{1\ov 2}\,x^{\alpha'\beta}x_{\alpha'\beta}$, 
the current becomes:
\be
j^{\langle ab\rangle }=  {R\ov 2(R^2+z^2)}
\pmatrix{ -dx^0\,\omega^{\alpha\beta}& -
dx^{\beta'\alpha} \cr
                                        dx^{\alpha'\beta}         &dx^0\,\omega^{\alpha'\beta'}}
\ee
and the action (\ref{eq:bosaction}) is  the conformally 
flat sigma model:
\be
{\cal L}_{S^5}=-j^{\langle{\bar a}{\bar b}\rangle}
j_{\langle{\bar a}{\bar b}\rangle}=
R^2{(dz)^2\ov (R^2+z^2)^2}~~.
\label{eq:s5metric}
\ee 
In vector notation the square of the vector $z$ should be consistent 
with the Wick-rotations 
that take us from $SO(5)$ to $SO(2,3)$. Indeed, we find that 
\be
(dz)^2= -(dx^0)^2-(dx^1)^2-(dx^3)^2+(dx^2)^2+(dx^4)^2
\ee 
which has the needed $(2,3)$ signature. 

One can arrive at the same result using only vector notation. In other 
words, one represents a 6-vector as a $4\times 4$ matrix using
the 5-dimensional $\gamma$ matrices and the charge conjugation 
matrix which can be chosen to be the $Sp(4)$-invariant metric:
\be
z^{ab}=x_6 \Omega^{ab}+ x^i (\Omega \gamma_i)^{ab}~~~~i=1,...,5~~~.
\label{eq:6vector}
\ee 
It is then straightforward to compute the (antisymmetric traceless 
part of the) current which turns out to be:
\be
j^{\langle ab\rangle }={1\ov x_6^2 + x^i x^j\eta_{ij}} x^{[6}\,dx^{i]}\,
(\Omega \gamma_i)^{ab} 
\ee
where $\eta$ is the $SO(2,3)$-invariant metric. As pointed out 
before, the $GL(1)$ gauge freedom 
can be used to fix a component of $x=(x^i, x^6)$ to any nonvanishing 
value. Choosing 
again to fix
the $GL(1)$ gauge by requiring that $x^6=R$, we obtain the 
$\sigma$-model in equation (\ref{eq:s5metric}). 
(Relations between various coordinates are discussed in the appendix.)

\newsection{Super}

In the previous section we constructed two  bosonic sigma models 
with target space the coset $GL(4)/(Sp(4)\otimes GL(1))$. They can 
be used to build sigma models with target space the Wick-rotated
$AdS_5\times S^5$. Now we proceed with the supercoset 
construction started in section 2 and define the currents
\be
{J_A}^B+{A_A}^B={Z_A}^M\,d{Z_M}^B~.
\label{eq:curr}
\ee
where ${A_A}^B$ is the $[Sp(4)\otimes GL(1)]^2$ connection 
and ${J_A}^B$ is the superspace analog of the antisymmetric traceless 
part of ${j_a}^b$ from the previous section.

As shown in \cite{berk2}, if in a coset $G/H$ the subgroup $H$ is 
the invariant locus of a particular ${\bf Z}_4$ automorphism of 
the group $G$, then the extra integral in the Wess-Zumino term 
can be performed explicitly and the result expressed  in terms of only  
components of the 
current (\ref{eq:curr}). In other words, under the above assumptions the 
bosonic components of the NS-NS 
super two-form vanish while the rest are constant \cite{CB}.  In our 
case $H=(Sp(4)\otimes GL(1))^2$, but 
only its $Sp(4)\otimes Sp(4)$ subgroup 
has the property mentioned above. This leads to slight deviations 
from the results of
\cite{berk2}. The action is given by: 
\be
S=\int_{\Sigma} J^{ab}\wedge *J_{ab} - J^{{\bar a}{\bar b}} \wedge  
*J_{{\bar a}{\bar b}} \pm
{1\ov 2}(E^{1/2} J^{a{\bar b}} \wedge J_{a{\bar b}} - 
E^{-1/2} J^{{\bar a}{b}} \wedge J_{{\bar a}{b}})
\label{eq:action}~~.
\ee
where $E={\rm sdet}\,{Z_M}^A$. The relative coefficient is 
not fixed by the coset construction. However,
the requirement of existence of $\kappa$ symmetry fixes its 
absolute value to be $1/2$. In the gauge
$E=1$ our coset reduces to $PSL(4|4)/Sp(4)\otimes Sp(4)$. However,
as we will see in section 7, other choices of $E$ can be useful as well.

Showing that the action (\ref{eq:action}) is $\kappa$-symmetric and 
finding the corresponding 
transformations is not a complicated task. Following the model of 
construction of the GS 
superstring in general supergravity background we define the variations
\be
{\Delta_A}^B = {Z_A}^M\delta {Z_M}^B~.
\ee

The $\kappa$ symmetry transformations that we consider 
have a form similar to the standard ones. In particular, the bosonic 
variations vanish
\be
\Delta_{ab}=0=\Delta_{{\bar a}{\bar b}}~.
\label{eq:kappa1}
\ee
This can be understood by recalling that the generator 
for $\kappa$ transformations
is $\sla{p}\,d$ where $d=0$ is the second-class constraint associated 
to the canonical 
momentum conjugate to the odd superspace coordinates 
and $p$ is the canonical 
momentum conjugate to $x$. By acting with it we get
\be
\delta\,.= [\sla{p}\,d\,,\,.\,]=[\sla{p}\,,\,.\,]d+ \sla{p}\,[d\,,\,.\,]~.
\ee
The first term vanishes by the second-class constraint while the 
second term contributes only fermionic variations.

The variations of the current $J$ can be obtained by splitting the 
naive variation of the right-hand-side
of equation (\ref{eq:curr}) in coset, $Sp(4)$ and $GL(1)$ parts and 
introducing the $(Sp(4)\otimes GL(1))^2$ covariant derivative ${\cal D}$ 
to absorb the $Sp(4)$ and $GL(1)$ pieces. With this 
provision it is then immediate to show that they are given by: 
\bea
\delta J_{ab}&=&{J_{\langle a}}^{\bar c}
\Delta_{{\bar c}|b\rangle }-{\Delta_{\langle a}}^{\bar c}
J_{{\bar c}|b\rangle }\nonumber\\
\delta J_{{\bar a}{\bar b}}&=&{J_{\langle {\bar a}}}^{c}
\Delta_{{c}|{\bar b}\rangle }
-{\Delta_{\langle {\bar a}}}^{c}J_{{c}|{\bar b}\rangle }\nonumber\\
\delta  J_{a{\bar b}}&=&{\cal D}\Delta_{a{\bar b}}+ 
{J_a}^c\Delta_{c{\bar b}}-
{\Delta_a}^{\bar c}J_{{\bar c}{\bar b}}\label{eq:genericvars}\\
\delta  J_{{\bar a}b}&=&{\cal D}\Delta_{{\bar a}{b}}+\
{J_{\bar a}}^{\bar c}\Delta_{{\bar c}{b}}-
{\Delta_{\bar a}}^{c}J_{{c}{b}}\nonumber
\eea
where we used the fact that $\delta E = 
E\,str\Delta={\Delta_a}^a-{\Delta_{\bar a}}^{\bar a}=0$ and
${\cal D}E=0$. In 
writing these variations we did not take into account the world 
sheet metric or, equivalently, the
world sheet zweibein ${e_i}^{\cal M}$. Using light-cone coordinates 
we define its variation as
\be
\delta {e_\pm}^{\cal M}=\Delta_{\pm +}{e_-}^{\cal M}+
\Delta_{\pm -}{e_+}^{\cal M}
\ee
which implies that the full variations of the current are
\be
\delta_\kappa {{J_{\pm}}_A}^B=\Delta_{\pm +}{{J_{-}}_A}^B + 
\Delta_{\pm -}{{J_{+}}_A}^B + \delta{{J_{\pm}}_A}^B   
\ee 
with the last term being given by (\ref{eq:genericvars}).

With this starting point it is straightforward to determine the 
fermionic variations 
$\Delta_{a{\bar b}}$ and $\Delta_{{\bar a}b}$
as well as the variations of the zweibein, $\Delta_{\pm\pm}$. We 
need to use only two of the Maurer-Cartan equations 
\bea
{\cal D} J_{a{\bar b}}&+& {J_a}^c\wedge J_{c{\bar b}}+
{J_a}^{\bar c}\wedge J_{{\bar c}{\bar b}}=0\nonumber\\
{\cal D}J_{{\bar a}{b}}&+&
{J_{\bar a}}^c\wedge J_{c{b}}+{J_{\bar a}}^{\bar c}\wedge 
J_{{\bar c}{b}}=0
\eea
together with the identities 
${J_{(i}}^{ab}{J_{j)}}_{ac}={1\ov 4}{\delta^b_c}{J_{(i}}^{ad}{J_{j)}}_{ad}$
(and similarly for all barred indices) 
coming from the $Sp(4)$ algebra. To make a long story short, the 
rest of the transformations that leave the action invariant are:
\bea
E^{1/4}\Delta_{a{\bar b}}&\pm&\sigma E^{-1/4}
\Delta_{{\bar b}a}={{J_\pm}_a}^c{\kappa}_{\mp c{\bar b}}+
{{J_\pm}_{\bar b}}^{\bar c}{\kappa}_{\mp a{\bar c}}\nonumber\\
\Delta_{+-}&+&\Delta_{-+}=0\label{eq:kappa2}\\
\Delta_{\pm\pm}&=&
-{1\ov 4} ( E^{-1/4}{{J_\pm}^{{\bar a} b}}\mp\sigma
E^{1/4}{{J_\pm}^{b {\bar a}}} ){\kappa}_{\pm b {\bar a}}\nonumber
\eea
where $\sigma=\pm 1$ matches the sign of the Wess-Zumino 
term in the action (\ref{eq:action})
and the $\kappa$ parameters do not transform under one of 
the local $GL(1)$ groups.
As we have already pointed out before, this computation is much 
easier than the one performed
in \cite{coset} because we do not need to decompose the currents 
in terms of coset generators.

\noindent
\newsection{Flat space limit}

The flat space limit of the model we constructed is taken in two steps:

1) Add the identity to a coset element and, according to dimensional 
analysis, divide its fermionic and bosonic components 
by $\sqrt{R}$ and $R$, respectively, and

2) expand for $R\rightarrow \infty$.

The first step implies that a  generic coset element is written as:
\be
{Z_M}^A={\delta_M}^A + {1\ov \sqrt{R}}{f_M}^A+{1\ov R}{b_M}^A 
\ee
Then, its inverse, as an expansion in  $1/R$, is given by:
\be
{Z_B}^N={\delta_B}^N - 
{1\ov \sqrt{R}}{\delta_B}^M{f_M}^D{\delta_D}^N
-{1\ov R}({\delta_B}^M{b_M}^D{\delta_D}^N-{\delta_B}^M
{f_M}^D{\delta_D}^P{f_P}^C{\delta_C}^N)+...
\label{eq:invZ}
\ee
Using these relations the current (\ref{eq:curr}) has the following 
expression:
\bea
{J_A}^B\approx {1\ov R}\Big[\!\!\!\!
&{}&\!\!{\delta_A}^M (d{b_M}^B-{f_M}^D
{\delta_D}^N d{f_N}^B)+\sqrt{R} {\delta_A}^M \,d{f_M}^B \\
&-&\!{1\ov \sqrt{R}}{\delta_A}^M({f_M}^D{\delta_D}^Nd{b_N}^B
+{b_M}^D{\delta_D}^Nd{f_N}^B - {f_M}^D{\delta_D}^P{f_P}^C
{\delta_C}^Nd{f_N}^B)\Big]+...
\nonumber
\eea


It is now immediate to separate the bosonic and fermionic components 
of the current. The even components  of the current to leading order 
in $1/R$ are given by
\be
\on\circ J_a{}^{b}={1\ov R}{\delta_a}^m (d{b_m}^b-
{f_m}^{\bar d}{\delta_{\bar d}}^{\bar n} d{f_{\bar n}}^b)
\label{eq:jab}
\ee
which resembles the currents from the usual GS superstring.
The expression for $\on\circ{J}_{{\bar a}}{}^{{\bar b}}$ is obtained by 
replacing unbarred indices with 
barred indices and vice-versa in the above equation. The odd 
components are given by:
\be
\on\circ{J}_{a}{}^{{\bar b}}={1\ov \sqrt{R}} {\delta_a}^m \,
d{f_m}^{\bar b}-
{1\ov R^{3/2}}{\delta_a}^m({f_m}^{\bar d}{\delta_{\bar d}}^{\bar n} 
d{b_{\bar n}}^{\bar b}
+{b_m}^d{\delta_d}^nd{f_n}^{\bar b} - {f_m}^{\bar d}
{\delta_{\bar d}}^{\bar p}{f_{\bar p}}^c
{\delta_c}^nd{f_n}^{\bar b})
\ee
and similarly for $\on\circ{J}_{{\bar a}}{}^{b}$. The current 
$J_{ab}$ appearing in the action (\ref{eq:action})
is actually only the traceless antisymmetric part of the r.h.s. of 
equation (\ref{eq:jab}), as required by the coset construction. Before
writing this explicitly, let us look for the field 
redefinitions that put the l.h.s. of the
$\kappa$-transformations in a form similar to the usual one. 

The $\kappa$ transformations in the flat space limit can be obtained from  
the equations (\ref{eq:kappa1}) and (\ref{eq:kappa2}) by making 
the rescaling $\kappa\rightarrow \sqrt{R}\kappa$, expanding 
around $R\rightarrow\infty$ 
and identifying the powers of $1/R$ on both sides of the equations. We get:
\be
\delta b_{ab} - {f_a}^{\bar d}\,\delta{f_{{\bar d}b}}=0~~~~~~~~
\delta b_{{\bar a}{\bar b}} - {f_{\bar a}}^d\,\delta{f_{d{\bar b}}}=0
\ee
from (\ref{eq:kappa1}) while from (\ref{eq:kappa2}) we get
\be
\delta (f_{a{\bar b}}\pm f_{{\bar b}a})=-R \,
\on\circ{J}_{\langle ac\rangle}
{{\kappa_{\pm}}^c}_{\bar b}
-R \,\on\circ{J}_{\langle {\bar b}{\bar c}\rangle }
 {{\kappa_{\pm }}_a}^{\bar c}~~.
\label{eq:flatkappa}
\ee

Since for the flat space GS superstring one usually 
writes $\delta\theta\sim\sla{\Pi}\kappa$,
the l.h.s. of transformations (\ref{eq:flatkappa}) suggest the 
following field redefinitions:
\be
\theta_{a{\bar b}}={1\ov 2}(f_{a{\bar b}}+ f_{{\bar b}a})~~~~~~~~~~~
\theta_{{\bar b}a}={1\ov 2}(f_{a{\bar b}}- f_{{\bar b}a})
\ee
In terms of these new objects the (traceless, antisymmetric part of the) 
bosonic currents become (we removed the overall factor of $1/R$):
\bea
\on\circ {J}_{ab}&=& dy_{\langle ab\rangle }-
({\theta_{\langle a}}^{\bar d} \,d\theta_{b\rangle {\bar d}} + 
{\theta}_{{\bar d}\langle a}\,d{\theta^{\bar d}}_{b\rangle })
~~~;~~~
y_{\langle ab\rangle }\equiv 
b_{\langle ab\rangle }+{\theta_{\langle a|}}^{\bar d}
{\theta_{{\bar d}|b\rangle }}
\nonumber\\
\on\circ{J}_{{\bar a}{\bar b}}&=&
dy_{\langle {\bar a}{\bar b}\rangle }-
({\theta_{\langle {\bar a}}}^{d} \,d\theta_{{\bar b}\rangle {d}} +
{\theta}_{d\langle {\bar a}}\,d{\theta^d}_{{\bar b}\rangle })~~~;~~~
y_{\langle {\bar a}{\bar b}\rangle }\equiv 
b_{\langle {\bar a}{\bar b}\rangle }+{\theta_{\langle {\bar a}|}}^{d}
{\theta_{{d}|{\bar b}\rangle }}
\label{eq:bbcurrent}
\eea
For these currents to resemble the
standard flat space ones,  we would like to 
identify $y_{\langle {a}{b}\rangle }$ and 
$y_{\langle {\bar a}{\bar b}\rangle }$ with the space-time 
coordinates written in $SO(5)\otimes SO(5)$ spinor 
notation. As counted before, the number of components 
matches, but we also have to make sure that 
the rest of the action has the right form. 

There are several contributions to the Wess-Zumino term. The leading 
one is $1/R\,df\wedge df$
but it is a total derivative and drops out. This is good, because the 
rest of the action is of the order $1/R^2$.
The remaining contributions come from the  cross terms in 
$\on\circ{J}{}^{a{\bar b}}\on\circ J_{a{\bar b}}$ and 
$\on\circ J{}^{{\bar b}a}\on\circ J_{{\bar b}a}$ together with a cross term 
between $1/R\,df\wedge df$ and a $1/R$ term
coming from the expansion of $E^{\pm1/2}$. In terms of 
the $\theta$ variables introduced above 
the Wess-Zumino term can be written, up to a total derivative, as 
\bea
{1\ov 4}[E^{1/2}\on\circ J{}^{{\bar b}a}\on\circ J_{{\bar b}a}-
E^{-1/2}\on\circ J{}^{a{\bar b}}\on\circ J_{a{\bar b}}] &=& 
d(b^{[{a}{b}]}+{\theta^{[{a}|{\bar d}}}{\theta_{\bar d}}^{|b]})\wedge
({\theta_{[a}}^{\bar d} \,d\theta_{b]{\bar d}} - {\theta}_{{\bar d}[a}\,
d{\theta^{\bar d}}_{b]})
\nonumber\\
&-&d(b^{[{\bar a}{\bar b}]}+
{\theta^{[{\bar a}|d}}{\theta_{d}}^{|{\bar b}]})\wedge
({\theta_{[{\bar a}}}^{d} \,d\theta_{{\bar b}]{d}} - 
{\theta}_{d[{\bar a}}\,d{\theta^d}_{{\bar b}]})
\nonumber\\
&+&{\theta}^{d[{\bar a}}\,d{\theta_{d}}^{{\bar b}]}\wedge 
{\theta_{[{\bar a}}}^{c}
\,d{\theta_{{\bar b}]c}}+ 
{\theta}^{{\bar d}[{a}}\,d{\theta_{\bar d}}^{{b}]}\wedge 
{\theta_{[{a}}}^{\bar c}
\,d{\theta_{{b}]{\bar c}}}+\nonumber\\
&+&d({b^a}_a-{b^{\bar a}}_{\bar a} + 2 \theta^{a{\bar b}}
\theta_{{\bar b}a})(\theta^{c{\bar d}}\,
d\theta_{c{\bar d}}+\theta^{{\bar d}c}\,d\theta_{{\bar d}c})
\eea
This is not the standard form of the Wess-Zumino term in flat 
space. However, we still need
to separate the traces out of every antisymmetric factor in the 
equation above. The contributions 
from the first two lines completely cancel the fourth line while 
in the third line the traces 
cancel due to the antisymmetry of the $\wedge$-product. Thus, 
with the definition of $y$ from 
equation (\ref{eq:bbcurrent}), the Wess-Zumino term becomes:
\bea
{1\ov 4}[E^{1/2}\on\circ J{}^{a{\bar b}}\on\circ J_{a{\bar b}} - 
E^{-1/2}\on\circ J{}^{{\bar b}a}\on\circ J_{{\bar b}a}]&=&
{\theta}^{d\langle {\bar a}}\,d{\theta_{d}}^{{\bar b}\rangle }\wedge 
{\theta_{\langle {\bar a}}}^{c}\,d{\theta_{{\bar b}\rangle c}}+ 
{\theta}^{{\bar d}\langle {a}}\,d{\theta_{\bar d}}^{{b}\rangle }\wedge
{\theta_{\langle {a}}}^{\bar c}\,d{\theta_{{b}
\rangle {\bar c}}}\nonumber\\
+dy^{\langle ab\rangle }\wedge
({\theta_{\langle a}}^{\bar d} \,
d\theta_{b\rangle {\bar d}} &-&
{\theta}_{{\bar d}\langle a}\,d{\theta^{\bar d}}_{b\rangle })-
d y^{\langle {\bar a}{\bar b}\rangle }\wedge
({\theta_{\langle {\bar a}}}^{d} \,d\theta_{{\bar b}\rangle {d}} - 
{\theta}_{d\langle {\bar a}}\,d{\theta^d}_{{\bar b}\rangle })~.
\eea
Combining this  with the $\on\circ J{}^{ab}\on\circ J_{ab}$ 
and $\on\circ J{}^{ {\bar a}{\bar b}}\on\circ J_{{\bar a}{\bar b}}$ terms
constructed from equations (\ref{eq:bbcurrent}) we get 
the usual Green-Schwarz action written 
in $SO(5)\otimes SO(5)$ spinor notation.

\newsection{Kallosh-Rahmfeld-Pesando gauge}

In the previous section we showed that the 
action (\ref{eq:action}) together with the $\kappa$ 
transformation rules (\ref{eq:kappa1}) and  (\ref{eq:kappa2}) 
reproduce, in the flat space limit, 
the  usual Green-Schwarz action. In this section we will find, for 
the curved space model, 
a $\kappa$-symmetry gauge that simplifies the action.  Since 
the action also has an
$(Sp(4)\otimes GL(1))^2$ local invariance, we need to fix it as 
well. In section \ref{bosonic} we constructed 
coset representatives of $GL(4)/Sp(4)\otimes GL(1)$ that 
reproduce the $AdS_5$ and $S^5$ sigma models.
With slight improvement they will continue to be a part of 
the supersymmetric construction.

There are many ways to parametrize 
the $GL(4|4)/(Sp(4)\otimes GL(1))^2$ coset representatives.
We will start with one that exhibits the $4+6$ splitting 
reminiscent of the D3 brane background. Schematically it looks as follows:
\be
Z=[x^{(4)}][\theta][x^{(6)}]
\label{eq:dec}
\ee
where  $[x^{(4)}]$ denotes the coordinates parallel  to the 
brane while $[x^{(6)}]$ describes the coordinates orthogonal 
to it. An advantage of this parametrization is that it produces a 
separation of the transformation of the various components. 
As far as the even generators of the (4-dimensional) 
superconformal group are concerned,  only the right-most factor 
transforms  under the $R$-symmetry group 
$SL(4)\subset SL(4|4)$. Ordinary supersymmetry transformations 
mix the  left-most factor with the middle one while the 
$S$-supersymmetry transformations mix all three factors together.

Using the $Sp(4)$ gauges introduced in 
section \ref{bosonic} and noting that the
coset representatives (\ref{eq:adsgauge}) giving the 
$AdS_5$ metric can be written as:
\be
{\cal  X}=\pmatrix{{\bf I}& 0\cr {\bf x}& x_0\,{\bf I}}=
\pmatrix{{\bf I}& 0\cr {\bf x}& {\bf I}}\,
\pmatrix{{\bf I}& 0\cr 0& x^0\,{\bf I}}\equiv {\bf X}\,{\bf X_0}
\ee
the explicit form of equation (\ref{eq:dec}) is given by
\be
Z=\pmatrix{ {{ X}_m}^d  &  0     \cr
                 0         & {\delta_{\bar m}}^{\bar n}  }
  \pmatrix{      {\delta_d}^c     &  {\theta_d}^{\bar p}     \cr
            {\theta_{\bar n}}^c    &     {\delta_{\bar n}}^{\bar p}     }
  \pmatrix{  {(X_0)_c}^b   &      0     \cr
                  0            &   {{z}_{\bar p}}^{\bar b}     }
\label{eq:cosetparam}
\ee
where we displayed the matrix indices to emphasize the 
transformation properties
of various blocks. In the above expression $z_{{\bar p}{\bar b}}$ is, 
for the time being, an
arbitrary $4\times 4$ antisymmetric matrix representing an 
arbitrary 6-vector 
which will describe the $S^5$ part of the space. We will reduce to five 
its number of independent components by fixing the gauge for 
the last $GL(1)$ factor. In section \ref{bosonic} we used the  
$GL(1)$  transformations to set to
$1$ the sixth component of ${\bf z}$ and we obtained the conformally 
flat metric for the sphere. 
As will become apparent shortly, this is not a convenient gauge 
in the supersymmetric context. 
For this reason we choose to fix the $GL(1)$ at the end.
This will further simplify the action.

Once the $Sp(4)$ gauges are fixed as above we are naturally 
led to pick a $\kappa$-symmetry gauge.
In general one can set to zero any component of the 
spinors $\theta$ which is  acted upon by
the $\kappa$ transformations. If we do not want to further 
break the global $SL(2)\otimes SL(2)$
invariance surviving after $Sp(4)$ gauge fixing, we are led 
to a fairly small number of choices: 
\bea
{\theta_{\alpha}}^{\bar n}=0~~~{\rm and}~~~
{\theta_{\bar m}}^{\alpha'}=0~~~~&;&~~~~
{\theta_{\alpha'}}^{\bar n}=0~~~{\rm and}~~~
{\theta_{\bar m}}^{\alpha}=0\nonumber\\
{\theta_{a}}^{\bar \mu}=0~~~{\rm and}~~~
{\theta_{\bar {\mu'}}}^{a}=0~~~~&;&~~~~
{\theta_{a}}^{\bar {\nu'}}=0~~~{\rm and}~~~
{\theta_{\bar \mu}}^{a}=0                                     \\
{\theta_{\alpha}}^{\bar n}=0~~~{\rm and}~~~
{\theta_{\bar m}}^{\alpha}=0~~~~&;&~~~~
{\theta_{\alpha'}}^{\bar n}=0~~~{\rm and}~~~
{\theta_{\bar m}}^{\alpha'}=0\nonumber
\eea
and, of course, linear combinations thereof. To get a 
simple action one needs that the inverse of the 
fermion matrix,  $[\theta]^{-1}$, has as few terms 
as possible. This narrows the possible choices to 
the first four listed above.  A closer look at the structure 
of the current reveals that the first set of gauge conditions 
\be
{\theta_{\alpha}}^{\bar n}=0~~~{\rm and}~~~
{\theta_{\bar m}}^{\alpha'}=0
\label{eq:gaugecond}
\ee
gives the simplest action. This is related to the fact 
that $[x^{(4)}]^{-1}\,d[x^{(4)}]$ has 
nonvanishing entries only in the block $(\alpha',\,\beta)$ and 
therefore has ``destructive 
interference'' with the fermion matrix. Furthermore, upon 
Wick-rotation back to $((4,1),(5,0))$ 
signature, the two parts of the gauge are complex conjugate to 
each other, which will lead to a real action.
This will be the gauge that we will consider in the following. 

At this point there exists the issue regarding the consistency of 
the gauge choice. By 
performing a $\kappa$ variation of the gauge conditions one 
can check that there is no left-over gauge invariance.  


The currents can be easily computed. We have used the 
gauge conditions and the previous observations 
to cancel various terms as well as to write the result with an 
apparent $Sp(4)\otimes Sp(4)$ symmetry:  
\bea
{J_{a}}^{b}&=&{(j_{AdS_5})_a}^b - {({X_0}^{-1})_a}^c \,
{\theta_c}^{\bar m}\,d{\theta_{\bar m}}^d\,
{{{X_0}}_d}^b\nonumber\\
{J_{\bar a}}^{\bar b}&=&{({z}^{-1})_{\bar a}}^{\bar m}\,
d{{z}_{\bar m}}^{\bar b}={(j_{S^5})_{\bar a}}^{\bar b}\\
{J_a}^{\bar b}&=& {({X_0}^{-1})_a}^c\,d{\theta_c}^{\bar m}\,
{{z}_{\bar m}}^{\bar b}\nonumber\\
{J_{\bar a}}^{b}&=&{{z}_{\bar a}}^{\bar m}\,d{\theta_{\bar m}}^{c}\,
{{( X_0)}_d}^b\nonumber
\eea
where ${(j_{AdS_5})_a}^b$ and ${(j_{S^5})_{\bar a}}^{\bar b}$ are 
the $AdS_5$ and $S^5$ bosonic currents, respectively.
It is now  straightforward to write down the various terms in the 
action. The $({\bar a}{\bar b})$ 
part of the current is the same as with no fermions. As mentioned 
before, ${\bf z}$ describes a 6-vector with
a scale invariance that remains to be fixed. However, we can 
already say that in the right coordinates
${J_{\bar a}}^{\bar b}$ produces the $S^5$ sigma model regardless 
of its norm. Indeed, as shown in the 
appendix, the coordinate transformation 
${z_m}\rightarrow Y^{mn}=z^{am}{z_a}^n$ 
allows us to express the metric on $S^5$ in terms of {\it only} the 
unit vector pointing along ${\bf Y}$.
Thus, regardless of how we chose to fix the norm of ${\bf Y}$, the 
$J^{{\bar a}{\bar b}}\,J_{{\bar a}{\bar b}}$ term in the action produces 
the standard metric on $S^5$. For this observation to be of any use 
we need the action to depend only on $Y$. As we will see shortly, 
this is indeed the case.

The $J^{ab}\,J_{ab}$ term in the action is equally easy. We get
\be
J^{ab}\,J_{ab}=(j_{AdS_5})^{\alpha\beta}\,(j_{AdS_5})_{\alpha\beta}+
(j_{AdS})^{\alpha'\beta'}\,(j_{AdS})_{\alpha'\beta'}+
2J^{\alpha'\beta}\,J_{\alpha'\beta}
\ee
where for the last term we used the antisymmetry of $J_{ab}$. The 
first two terms are equal and given by
\be
(j_{AdS})^{\alpha\beta}\,(j_{AdS})_{\alpha\beta}=
(j_{AdS})^{\alpha'\beta'}\,(j_{AdS})_{\alpha'\beta'}=
{1\ov 2}\left({dx^0\ov x^0}\right)^2
\ee
as follows immediately from equation (\ref{eq:jads}).  The last 
term, which also receives 
fermionic contributions, has the expression
\be
2J^{\alpha'\beta}\,J_{\alpha'\beta}={1\ov 2\,(x^0)^2}\
\,(dx^{\alpha'\beta}-{\theta}^{\alpha'{\bar m}}\,
d{\theta_{\bar m}}^\beta)
(dx_{\alpha'\beta}-{\theta_{\alpha'}}^{\bar m}\,
d{\theta_{{\bar m}\beta}})~~.
\ee 
The way it stands this term is not real upon Wick-rotating back to signature
$(1,9)$. However, this can be problem can be solved using some information 
from the flat space limit. There we were naturally led to redefine
the bosonic coordinates by absorbing a $({\rm fermion})^2$ piece and thus
putting the coordinates in a chiral-like representation of supersymmetry.  
In the present situation, by redefining 
\be
x^{\alpha'\beta}\rightarrow x^{\alpha'\beta}+{1\ov 2} \theta^{\alpha' {\bar m}}
\theta_{\bar m}{}^\beta
\ee
each bracket becomes 
\be
dx^{\alpha'\beta}+{1\ov 2} (d\theta^{\alpha' {\bar m}}
\theta_{\bar m}{}^\beta)-{1\ov 2} (\theta^{\alpha' {\bar m}}
d\theta_{\bar m}{}^\beta)
\ee
and the two $({\rm fermion})^2$ terms are, after Wick-rotation, 
conjugate to each-other. 

As mentioned before, the structure of the Wess-Zumino term 
will decide whether
the action can indeed be expressed only in terms of ${\bf Y}$. Using 
its definition, the first part of the Wess-Zumino term is given by
\be
E^{-1/2}J^{{\bar a}b}J_{{\bar a}b}= {1\ov x^0\,|{\bf Y}|}
{Y}^{{\bar m}{\bar n}}\,d{\theta_{\bar m}}^\alpha \wedge\,
d\theta_{{\bar n}\alpha}~~,
\label{eq:wz1}
\ee
while the second part takes the form
\be
E^{1/2}J^{a{\bar b}}J_{a{\bar b}}=-{|{\bf Y}|\ov x^0}
{{Y}^{-1}}_{{\bar m}{\bar n}}\,
d{\theta}^{\alpha'{\bar m}} \wedge\,
d{\theta_{\alpha'}}^{\bar n}~~.
\label{eq:wz2}
\ee
The sign in the second equation comes from the fact that while 
${Y}^{{\bar m}{\bar n}}=z^{{\bar a}{\bar m}}{z_{\bar a}}^{\bar n}$, 
its inverse is ${{Y}^{-1}}_{{\bar m}{\bar n}}=
-{z_{\bar m}}^{\bar a}z_{{\bar n}{\bar a}}$ and 
$|{\bf Y}|^2=1/8\,\epsilon_{{\bar m}{\bar n}{\bar p}{\bar q}}
Y^{{\bar m}{\bar n}} Y^{{\bar  p}{\bar q}}$.  We get therefore 
that the action can 
be expressed in terms of only ${\bf Y}$ and its inverse. This  in turn 
implies that we are free to choose its norm without altering the 
form of the action.

We argued before that with the gauge fixing considered here the 
action will be real upon Wick-rotating back to $((4,1),(5,0))$ 
signature. This might not be obvious from equations
(\ref{eq:wz1}) and (\ref{eq:wz2}). To see this we first notice that
using the (trivial) identity $\delta^i_{[a}\epsilon^{}_{bcde]}=0$, 
${\bf Y}^{-1}$ is proportional to ${\bf Y}$: 
\be
{{Y}^{-1}}_{{\bar m}{\bar n}}=-{1\ov 2|{\bf Y}|^2}
\epsilon_{{\bar m}{\bar n}{\bar p}{\bar q}}{{Y}}^{{\bar p}{\bar q}}~~.
\label{eq:inverse}
\ee
Under these circumstances the Wess-Zumino term becomes real 
since using the $SU(4)\simeq SO(6)$ algebra it can be shown 
that after Wick rotation 
\be
{{Y}^\dagger}_{{\bar m}{\bar n}}=-{1\ov 2}
\epsilon_{{\bar m}{\bar n}{\bar p}{\bar q}}{{Y}}^{{\bar p}{\bar q}}
\ee
where we used the fact that the norm of ${\bf Y}$ is real.

Recall that we are still to fix the last $GL(1)$ gauge. From 
the equations (\ref{eq:wz1}-12) it is clear that the most useful gauge is
\be
|{\bf Y}|={1\ov x^0}~~~\Leftrightarrow~~~E=1
\label{eq:lastgauge}
\ee
We therefore expect that the resulting action will
be equivalent to the one obtained in \cite{gaugefixing}. This will 
indeed be the case.

As pointed out before, the Wess-Zumino term allows the action to be  
naturally written in terms of ${\bf Y}$. As shown in 
the appendix, the bosonic sigma model depends only on the unit vector 
along ${\bf Y}$.
Thus, with the gauge (\ref{eq:lastgauge}) we get upon Wick-rotation 
the metric on the unit 5-sphere
\be
J^{{\bar a}{\bar b}}J_{{\bar a}{\bar b}}=(d\Omega_5)^2
\ee 
which together with the 
$(j_{AdS})^{\alpha\beta}\,(j_{AdS})_{\alpha\beta}+
(j_{AdS})^{\alpha'\beta'}\,(j_{AdS})_{\alpha'\beta'}=
\left({dx^0\ov x^0}\right)^2$ combine and give just
\be
{(d{\bf Y})^2\ov {\bf Y}^2}
\ee
i.e.~a conformally flat 6-dimensional space in cartesian coordinates.

Collecting various partial results from  this section 
we find the $\kappa$ gauge fixed action to be
\bea        
{\cal S}=\int_\Sigma\,{1\ov 2\,(x^0)^2}\
\,(dx^{\alpha'\beta}&+&{1\ov 2}[d{\theta}^{\alpha'{\bar m}}\,
{\theta_{\bar m}}^\beta-{\theta}^{\alpha'{\bar m}}\,
d{\theta_{\bar m}}^\beta])^2
+{1\ov {\bf Y}^2}(d{\bf Y})\wedge *(d{\bf Y})\nonumber\\
&+&{1\ov 2}\left[ \,{\bf Y}^{{\bar m}{\bar n}}\,
d{\theta_{\bar m}}^\alpha \wedge\,
d\theta_{{\bar n}\alpha}-{1\ov 2}
\epsilon_{{\bar m}{\bar n}{\bar p}{\bar q}}
{{\bf Y}}^{{\bar m}{\bar n}}
d{\theta}^{\alpha'{\bar p}} \wedge\,d{\theta_{\alpha'}}^{\bar q}\,\right]~~
\label{eq:gfaction}
\eea
where the square is taken with $\wedge *$.
This action has a form equivalent to the 
one derived in \cite{gfkalloshramfeld}. There and in \cite{tskall} it 
has been checked that the fermionic quadratic form has 
no left-over zero modes around inhomogeneous
($\sigma$-dependent) string configurations. Furthermore, this 
gauge can be reached from any point
in the space of such configurations.

\newsection{Complex gauge}

In deriving the equation (\ref{eq:gfaction}) we have been guided by the 
requirement that the action be hermitian after 
we Wick-rotate back to the original coset superspace. If, however, we 
relax this assumption, we can 
find gauge conditions which bring the action to  an even simpler 
form. Such an example is the gauge
\be
{\theta_{\bar m}}^{n}=0
\ee
together with the coset parametrization 
\be
Z= \pmatrix{      {{\delta}_m}^n     &  {\theta_m}^{\bar n}     \cr
            {\theta_{\bar m}}^n    &     {{\delta}_{\bar m}}^{\bar n}     }
  \pmatrix{  {{x}_n}^b   &      0     \cr
                  0            &   {{z}_{\bar p}}^{\bar b}     }~~.
\label{eq:newparam}
\ee
For the time being we did not fix any of the local gauge 
invariances.  These assumptions 
lead to the following set of currents:
\bea
{J_{a}}^{b}&=&{({x}^{-1})_{ a}}^{ m}\,d{{ x}_{ m}}^{ b}=
{(j_{AdS})_a}^b\nonumber\\
{J_{\bar a}}^{\bar b}&=&{{z}^{-1}_{\bar a}}^{\bar m}\,
d{{z}_{\bar m}}^{\bar b}=
{(j_{S^5})_{\bar a}}^{\bar b}\\
{J_a}^{\bar b}&=& {{{x}^{-1}}_a}^n\,d{\theta_n}^{\bar m}\,
{{z}_{\bar m}}^{\bar b}~~~~~~
~~~~~{J_{\bar a}}^{b}=0~~.\nonumber
\eea
For later convenience we define, as in the previous 
section, ${\bf Y}$ and the corresponding object 
for $AdS_5$,  say ${\bf W}$, as
\be
Y^{{\bar m}{\bar n}}=z^{{\bar a}{\bar m}}
{z_{\bar a}}^{\bar n}~~~~~~~ {W}^{mn}=x^{am}{x_a}^n
\ee
in terms of which the superdeterminant  is 
just $E=|{\bf Y}|^2/|{\bf W}|^2$. In terms of 
these objects the Wess-Zumino term is now:
\be
E^{1/2}
J^{a{\bar b}}J_{a{\bar b}}
=-{|{\bf Y}|\ov |{\bf W}|}{W}^{np}d{\theta_n}^{\bar m}\wedge 
d{\theta_p}^{\bar s}\,{{Y}^{-1}}_{{\bar m}{\bar s}}
=-{1\ov 2|{\bf W}||{\bf Y}|}\epsilon_{{\bar r}{\bar t}{\bar m}{\bar s}}
{W}^{np}{Y}^{{\bar r}{\bar t}}d{\theta_n}^{\bar m}
\wedge d{\theta_p}^{\bar n}~~.
\ee
We notice that without fixing any gauge, the 
Wess-Zumino term is expressed in terms
of {\it only} the unit vectors pointing in the 
direction of ${\bf W}$ and ${\bf Y}$. This feature can be
extended to the other terms as well. Choosing the $Sp(4)$ gauges 
for both $x$ and $z$ as in 
equation (\ref{eq:s5gauge}), the $AdS_5$ and $S^5$ metrics are 
the scale invariant ones when 
expressed in terms of ${\bf W}$ and ${\bf Y}$, as shown in 
the appendix. We therefore write 
the action as
\be
{\cal S}=\int_\Sigma\,|d{\bf W}|^2-|d{\bf Y}|^2+
{1\ov 4}\epsilon_{{\bar r}{\bar t}{\bar m}{\bar s}}
{W}^{np}{Y}^{{\bar r}{\bar t}}d{\theta_n}^{\bar m}
\wedge d{\theta_p}^{\bar n}
\ee
with the provision that ${\bf W}$ and ${\bf Y}$ represent 
unit 6-vectors. Let us emphasize that
this is not the result of $GL(1)^2$ gauge fixing. As pointed out 
before, this action is not hermitian 
any more when Wick-rotated back to the original superspace. 
Its hermitian conjugate is the action 
obtained with the gauge fixing condition
\be
{\theta_{m}}^{\bar n}=0
\ee
and the same parametrization of the coset elements 
as in (\ref{eq:newparam}).

\newsection{Conclusions}

In this paper we have followed a path different from the usual 
supercoset construction of the GS action 
on $AdS_5\times S^5$. By Wick rotations and Lie algebra
identifications we brought the coset to 
$GL(4|4)/(Sp(4)\otimes GL(1))^2$.  This modified 
starting point leads to a number of simplifications: 

-unconstrained matrices are used  instead of exponential 
parametrization of coset elements

-spinor notation is natural and leads to the elimination of  Dirac matrices 
and their identities

-the derivation of the action is more streamlined

-an easier proof of $\kappa$ invariance

-the flat space limit can be taken explicitly, without 
the use of {\it a priori} 
knowledge of $AdS_5\times S^5$ metric

-$\kappa$ gauge fixing is more transparent; the 
Kallosh-Rahmfeld-Pesando gauge is easily obtained based 
on reality and conformal invariance requirements

-the use of complex gauges is easier in our approach and it leads to 
simpler actions 
than previously considered

In \cite{conformal} it has been shown that the sigma model on 
the $PSL(n|n)$ 
supergroup manifold is exactly conformal. Since the $Sp(4)$ sigma 
models are 
also conformal, the $GL(1)$'s are abelian groups and our construction 
is GKO-like, we draw the conclusion that it is not unlikely that our 
construction leads to a conformal field theory. 

\bigskip

\acknowledgments

This work was supported by NSF grant $\# 9722101$.

\setcounter{section}{0}
\setcounter{subsection}{0}

\appendix{}

Using the parametrization of 6-vectors written in (\ref{eq:6vector}) it
is easy to see that 
the interpretation of $w^{mn}=z^{am}{z_a}^n$ as space-time coordinates 
is, before $GL(1)$ gauge fixing, nothing more that 
the coordinate transformation to a {\it manifestly} scale-invariant
metric. This 
statement is actually independent of the dimension.

We start from an antisymmetric matrix ${\bf z}$ and define ${\bf w}$ as:
\be
w^{ab}=z^{ca}\Omega_{cd} z^{db}
\ee
Decomposing this equation in vector notation provides us with the
relation between 
the D-vector $w^I$ and  the D-vector $z^I$. It is:
\be
w^0=(z^0)^2-(z^i)^2~~~~~~~w^i=2\, z^0\, z^i
\label{eq:tw}
\ee
where, as before, the indices $i$ and $j$ are $D-1$-dimensional indices.

Now starting from 
\be
ds^2=\left(d{w\ov \sqrt{w^2}}\right)^2=
{(dw^I)^2\ov (w^I)^2}-{(w^I\,dw_I)^2\ov (w^I)^4}
\label{eq:sim}
\ee
and using the change of variables (\ref{eq:tw}) we get the following 
line element:
\be
ds^2=4{(z^i \,dz^0 - z^0\,dz^i)^2\ov ((z^0)^2 + (z^i)^2)^2}
\ee
i.e.~a conformally flat metric in the gauge $z^0=1$. 

This transformation is part of a one-parameter family of 
transformations which 
for particular values gives D-dimensional versions of the 
orthographic, stereographic and gnomonic projections of 
the complex plane. 

One usually considers the projection of a sphere onto a plane 
tangent to the sphere. Here we implement 
this operation through a gauge condition. We start with the 
metric (\ref{eq:sim}) and let the plane 
pass through the point $p=(w^0, 0,...,0)$ while the sphere has 
radius $R=\sqrt{(w^I)^2}=\sqrt{(w^0)^2+(w^i)^2}$.  Instead 
of considering a sphere of fixed radius and perform a projection  
from an arbitrary point $q$  on the line linking the 
center of the sphere and the point $p$, we let the sphere expand 
and require that the distance between 
$q$ and $p$ be fixed. Considering $q$ at distance $aR$ from 
the center of the sphere and noticing that
the distance between the center of the sphere 
and $p$ is just $P=w^0$ we have the condition 
\be
P+aR=1+a
\ee
For various values for $a$ one recovers well-known coordinate systems:

$\bullet~~a=0~$   $\Rightarrow~$  $w^0=1$ and the metric becomes:
\be
ds^2={(dw^i)^2\ov 1+(w^i)^2}-{(w^i\,dw_i)^2\ov (1+(w^i)^2)^2}
\ee
i.e.~gnomonic projection.

$\bullet~~a=1~$ $\Rightarrow~$  $w^0=1-{1\ov 4}(w^i)^2$ and 
the metric becomes:
\be
ds^2={(dw^i)^2\ov (1+{1\ov 4}(w^i)^2)^2}
\ee
i.e.~stereographic projection

$\bullet~~a=\infty~$ $\Rightarrow~$  $w^0=\sqrt{1-(w^i)^2}$
\be
ds^2=(dw)^2~~~{\rm with}~~{(w^0)^2+(w^i)^2=1}
\ee
i.e.~orthographic projection

Using the coordinate transformations (\ref{eq:tw}) it is easy to 
translate these conditions in terms of $z$. The gauge 
condition $z^0=1$ together with the rescaling 
$z^i\rightarrow z^i/2$ reduces the equation (\ref{eq:tw})  to 
the case $a=1$.

\end{document}